# On-chip Amorphous THz Topological Photonic Interconnects

Rimi Banerjee[1,#], Abhishek Kumar[1,#], Thomas Tan Caiwei[1], Manoj Gupta[1], Ridong Jia[1], Pascal Szriftgiser[2], Guillaume Ducournau[3], Yidong Chong[1,4,*], and Ranjan Singh[1,4,*]

[1]Division of Physics and Applied Physics, School of Physical and Mathematical Sciences, Nanyang Technological University, 21 Nanyang Link, Singapore 637371, Singapore.

[2]Laboratoire de Physique des Lasers, Atomes et Molécules, PhLAM, UMR 8523 Université de Lille, CNRS, 59655 Villeneuve d'Ascq, France.

[3]Univ. Lille, CNRS, Centrale Lille, Univ. Polytechnique Hauts-deFrance, UMR 8520 - IEMN-Institut d'Electronique de Microélectronique et de Nanotechnologie, F-59000 Lille, France.

[4]Centre for Disruptive Photonic Technologies, The Photonics Institute, Nanyang Technological University, 50 Nanyang Avenue, Singapore 639798, Singapore.

[#]These authors contributed equally to this work.

[*]E-mail: yidong@ntu.edu.sg (Y.C.); ranjans@ntu.edu.sg (R.S.)

**Abstract**

Valley Hall photonic crystals (VPCs) offer the potential to create topological waveguides capable of guiding light through sharp bends on a chip. They can seamlessly integrate with functional components while occupying minimal space, making them a promising technology for terahertz (THz) topological photonic integrated circuits. However, a significant limitation for THz-scale integrated VPC-based devices has been the absence of arbitrary bend interconnects. Due to the crystalline symmetry, the traditional VPC designs restrict waveguides to the principal lattice axes (i.e., only 0-, 60- or 120- degree orientations). Here, we present an on-chip, all silicon implementation of deformed VPCs enabling topological waveguides with a variety of shapes and bends. Although the lattice is amorphous and lacks long-range periodicity, the topological protection of the waveguides is sustained by short-range order. We experimentally demonstrate the robust on-chip transmission of THz waves through waveguides of complicated shapes and arbitrary bends. We implement an amorphous lattice that serves as a frequency-dependent router capable of splitting the input signal into two perpendicular output ports, which cannot be achieved with an undeformed VPC. In addition, we showcase on-chip THz communication through $90^o$ and P-shaped VPC waveguides, achieving data rates of 72 Gbit/s and 32 Gbit/s, respectively. Our findings demonstrate that the amorphous topological photonic crystals significantly enhance the adaptability of on-chip interconnections while preserving the performance of the topological waveguides.

**Introduction**

The terahertz frequency domain (0.1-10 THz) has attracted significant research interest due to its broad range of applications in biological imaging[1], security scanning[2], and communication[3-7]. Notably, this frequency range is an excellent choice for 6G wireless communications[6,8-9]. However, many existing THz device components, including hollow metallic waveguides[10], metallic transmission lines[11], are bulky, lossy, and have high manufacturing cost and complexity. These shortcomings can be addressed with silicon (Si) based photonic crystals (PCs) [12-13], which can be used to implement low-loss terahertz waveguides[14-15] compatible with Complementary Metal Oxide Semiconductor (CMOS) technology, allowing for easier integration with various active and passive components to achieve advanced functionalities[15]. However, traditional silicon photonic circuit designs face certain constraints, including limited bandwidth and the emergence of significant transmission losses when waveguides are subjected to excessively sharp bends, thereby reducing the operational bandwidth. Suppressing bending losses is important for the integration of different functional components into compact, multi-component integrated THz devices[12-13].

A solution that has recently emerged is to use valley Hall photonic crystals (VPCs), which host topologically protected edge states at interfaces between domains with distinct topological band invariants (valley Chern numbers)[16-22]. Compared to other topological lattices, VPCs are relatively easy to implement on-chip, simply by breaking the spatial inversion symmetry of an underlying photonic structure, without breaking time-reversal symmetry (TRS)[21,23,24]. The topological edge states have linear dispersion relations and lie below the light line, which suppresses out-of-plane radiation losses. The edge states can maintain robust transport even when subjected to sharp bends of $60^o/120^o$, with transmittance and bandwidth comparable to an unbent waveguide[25-29]. Typically, conventional waveguides lacking this "topological protection" suffer extreme losses at comparably sharp bends[29-32]. The first experimental study of a THz VPC waveguide, by Yang *et al.*, found near-unity transmission over multiple consecutive bends, including a sample with five $120^o$ and five $60^o$ bends, and demonstrated on-chip communication with a data rate of 11 Gbit/s[21]. Subsequent refinements have been made to this VPC platform, including progress on the integration of the topological waveguides with components such as beam splitters[33], switches[34], high-Q topological cavities[35], sensors[36], and multiplexers[37]. It is also worth noting that in the THz regime, Anderson localisation induced by fabrication disorder is negligible, as such disorder typically has length scale much smaller than the operating wavelength (at higher frequencies, fabrication disorder may be a significant concern for VPC waveguides as well as other PC waveguides)[38,39].

To realize compact multi-component THz devices, it is desirable to have freeform waveguides that can serve as efficient interconnects between functional components at arbitrary locations. VPC waveguides appear to be limited in this respect, as they must run along one of the three principal axes of the VPC and can thus only bend by $60^o$ or $120^o$[29,33,40]. VPC waveguides composed of sequential $60^o$ and $120^o$ bends[17,18] do not constitute fully freeform waveguides.

Motivated by non-periodic lattices, such as quasicrystals and amorphous lattices[41-45], we have designed and implemented arbitrarily shaped topological waveguides based on controlled deformations of a VPC. The amorphous VPC is generated with the aid of a molecular dynamics simulator[46], which close-packs structural motifs in a locally graphene-like configuration, analogous to a crystalline VPC but without long-range crystalline order. As with previously realised VPCs, the design is all-dielectric and does not require TRS breaking. The lattice deformations reduce but do not eliminate the bulk bandgap; the internal domain walls of amorphous VPC can serve as waveguides based on the same topological principles as in crystalline VPCs. We fabricate these amorphous VPC waveguides on an all-silicon platform, and experimentally demonstrate robust transmission under various bends ($70^o$, $90^o$, and $105^o$). We also characterize a VPC waveguide with a more complex "P" shape. To illustrate the usefulness of this approach for functional THz devices, we design a frequency dependent router and splitter which can efficiently deliver an input signal into two output waveguides oriented perpendicular to the input waveguide. We present numerical comparisons to conventional (non-topological) waveguides, showing that amorphous VPCs provide more robust transmission across a wide operating band. To our knowledge, amorphous VPCs have not previously been studied in THz photonics, or in an all-dielectric on-chip platform. Our work thus expands the possibilities for the use of topological waveguides in functional THz devices.

## Results

We fabricate THz-scale amorphous VPCs like the one shown in Fig. 1**a**, which hosts a waveguide that transmits a THz signal from an input port to a perpendicularly oriented output port. The amorphous VPC consists of air holes etched into a high-resistivity silicon (10 kΩ.cm) wafer of 200 $\mu$m thickness and 11.7 relative permittivity. The structure in Fig. 1**a** comprises two domains, analogous to the two topologically distinct phases of a crystalline VPC[20,21,27]; the boundary between the two domains is indicated by a red line. Each domain consists of a hexagonal lattice with two sites per unit cell, indicated in inset (i) of Fig. 1**a**. In one domain, the unit cell (blue) contains an equilateral triangular air-hole and an inverted triangular air-hole, of side-lengths $l_1$ and $l_2$ respectively; in the other domain, the side-lengths are interchanged (magenta). When $l_1 = l_2$, the transverse electric (TE) band structure of the crystalline VPC hosts a pair of degenerate Dirac points at the high symmetry points $K$ and $K'$, protected by TRS and inversion symmetry (Fig. 1**b**, green curve). By taking $l_1 \neq l_2$, we break the inversion symmetry and lift the Dirac point degeneracy, opening a bandgap (Fig. 1**b**, violet curve). The two valleys at $K$ and $K'$ have valley Chern numbers with opposite signs (Fig. 1**c** and Supplementary Information S1); these signs flip in the opposite domain, i.e., when we interchange $l_1$ and $l_2$ (see Supplementary Information S1), so the two domains are topologically distinct[16]. At the domain wall, there should hence be topologically protected waveguide modes at frequencies corresponding to the bulk bandgap[16-18]. This edge state is highly localised and exhibits valley-locked robust transport, with opposite propagation directions in the two valleys (see Supplementary Information S1).

By varying the positions and orientations of the triangular air-holes (with side lengths $l_1$ and $l_2$ fixed), we smoothly deform the VPC so that its domain wall lies along a desired path—in this case, one featuring a $90^o$ bend between the input and output ports (Fig. 1**a**). The deformations are generated by using the molecular dynamics simulator LAMMPS[46] to close pack an auxiliary triangular lattice[47] (see Supplementary Information S2). Notably, the structure maps *locally* to a crystalline VPC with an average lattice periodicity of 245 $\mu$m but lacks true long-range periodicity. In the vicinity of the domain wall, the structure is locally equivalent to a domain wall between two VPCs with opposite valley Chern numbers. In Fig. 1**d**, we plot the experimentally recorded transmission spectrum, showing nearly unity transmission within a bandwidth of approximately 21

GHz (-3 dB window relative to the highest transmission value, shaded in yellow). We also numerically calculate the transmission, which agrees well with the measurement (see Supplementary Information S3). To characterize the dispersion of the waveguide, we calculate the group delay and observe that it is almost constant in the frequency range corresponding to the bulk bandgap (see Supplementary Information S3). In our experiment, the transmission is measured using a vector network analyzer (VNA) based setup (see Supplementary Information S4). A metallic hollow rectangular waveguide (WR3.4/WR2.2) attached to the VNA extender module is used to generate frequencies in the desired range. Before recording the signal through the sample, the VNA testbed is calibrated using the TRM (Through/Reflect/Match) method.

In Fig. 1**e**, we show the results of numerical simulations for a source antenna and probe antenna placed at different positions (marked respectively by stars and filled circles in Fig. 1**a**) away from the domain wall; the normalized field distribution ($|H_z|$) shows a dip at around 0.33 THz, consistent with the persistence of the bulk bandgap. In Fig. 1**f**, we plot the simulated field intensity distribution ($|H_z|^2$) at frequency 0.328 THz, obtained by a 3D full-wave simulation (see Methods and Supplementary Information S3); the resulting intensity distribution is strongly localized along the domain wall, as expected of a VPC edge state. For comparison with conventional (non-topological) waveguiding, we numerically investigate a line defect waveguide formed by removing a single row along the desired route on a single-domain VPC with identical lattice parameters; the simulation shows that most of the THz wave is blocked at the waveguide bend, resulting in a highly non-uniform intensity distribution along the path (see Supplementary Information S5). The topological waveguide, however, experiences negligible backscattering at the bend and exhibits a longitudinally uniform intensity distribution (Fig. 1**f**), like previously studied crystalline VPCs with $60^o$ or $120^o$ waveguide bends[29-32].

We additionally fabricated two samples with bends of $70^o$ (Fig. 2**a**) and $105^o$ (Fig. 2**b**). The experimentally measured transmission data is plotted in Fig. 2**c** (blue and red curves, respectively). Compared to the $70^o$ case, the bandwidth for the $105^o$ case is reduced by approximately 7 GHz and the transmission is 3 dB lower. The simulated field intensity distributions corresponding to $70^o$ and $105^o$ bends are shown in Fig. 2**d**, showing effective routing around the bends with negligible reflection. We also performed numerical transmission studies for amorphous VPC configurations with different bends, and the results consistently reveal a high transmission rate (see Supplementary

Information S6).

Fig. 2**e** shows a photograph of a more intricate structure with a waveguide forming a "P" shape, incorporating three sharp bends and curved paths. Zoomed-in views of the fabricated sample are shown in Supplementary Information S7. The experimentally measured transmission is plotted in Fig. 2**c** (purple curve). The waveguide still operates well, with a slightly reduced bandwidth and around 10 dB lower transmittance. The simulated field distribution ($|H_z|$) at 0.32 THz is plotted in Fig. 2**f**. In Supplementary Information S5, we numerically compare the transmission of this amorphous VPC waveguide with a conventional (nontopological) waveguide based on a line PC defect. The latter suffers from much higher loss (around 20 dB lower transmittance on average) due to localization as well as back-reflection at the bends[47].

We can leverage the amorphous VPC to design functional THz devices. Fig. 3**a** shows an example of a four-port frequency-dependent router, featuring two output ports (Port 3 and Port 4) oriented perpendicular to the input port, a situation that cannot be achieved with an undeformed VPC. There is an evanescent coupling between the input waveguide (connecting ports 1 and 2) and output ports 3 and 4, whose coupling strength depends on the separation ($D$) between the input/output waveguides and the common length ($L$) between them. Zoomed-in views of the sample are shown in Supplementary Information S7. The experimentally measured transmission is presented in Fig. 3**b.** In the 0.327-0.334 THz range (shaded in yellow), the signal is routed almost entirely to ports 3 and 4, consistent with the numerically simulated field intensity distribution shown in Fig. 3**c.** Transmission to port 2 is heavily suppressed (around -33 dB averaged over the operating range). The 8 dB discrepancy in transmission measured at ports 3 and 4 is mainly due to mismatched tapered couplers inserted at these ports (see insets (ii) and (iii) of Fig. 3**a**). We also performed numerical studies showing that the bandwidth over which this re-routing occurs can be further increased by decreasing the separation $D$ (see Supplementary Information S8). For frequencies outside this range but still within the bulk bandgap (cyan shaded area), most of the signal is transmitted to port 2 (average transmission is around -6 dB) with negligible transmission to ports 3 and 4 (around -26 and -34 dB, respectively). This is consistent with the simulated field intensity distribution shown in Fig. 3**d**. In short, this device serves as a demultiplexer enabling multiple functionalities including splitting and routing.

We conducted a high-speed data communication (datacom) experiment to demonstrate the practical use of the amorphous VPCs as an interconnect chip. Figs. **4a,d** depict the experimental setup. Adiabatic tapers are used to in-couple and out-couple the data signals to $90^o$ and "P"-shape VPC interconnect chips. The datacom measurements are carried out using a Uni-Carrier-Travelling Photodiode (UTC-PD) based THz communication setup (see Supplementary Information S4), employing quadrature amplitude modulation (QAM-16) to achieve a high data rate. Figs. **4b,e** shows the I-Q constellation diagrams corresponding to 18 GB (72 Gbit/s) and 8 GB (32 Gbit/s), respectively, recorded through $90^o$ and "P"-shape VPC chips at 331 GHz carrier frequency. The maximal baudrate was fixed to meet the available bandwidth of each device. The well isolated constellation points (Figs. **4b,e**) indicate minimal distortion in amplitude and phase of the transmitted signal, affirming the robust transport of data signals through VPC chips.

To assess the performance of high-speed data communication, we plotted the error vector magnitude (EVM) against relative transmission power in Figs. **4c,f**. The EVM is a performance indicator that quantifies the quality of the received datacom signals, representing the difference between the ideal transmitted symbols (without the chip) and the symbols recovered through the VPC chips. The red and black dots in Figs. **4c,f** represent the EVM data points with and without the VPC chip, respectively. Evidently, increasing the transmitted power improves the signal-to-noise ratio (SNR), resulting in reduced the EVM for data transmission for both the with and without chip. The introduction of the VPC chip is accompanied by an expected power penalty, indicated by a shift between the reference and VPC chip points, due to the additional transmission power required to overcome the insertion losses. The impact of the VPC does not exceed the expected penalty induced by the transmission losses previously measured using the VNA testbed, validating the use of this type of interconnect to handle wide band signal to carry data for chip-to-chip communications.

**Conclusions**

We have implemented topologically protected THz waveguides of arbitrary shape, using amorphous VPCs on a silicon on-chip platform without TRS breaking. These waveguides exhibit high transmission, and topological protection against backscattering akin to crystalline VPC waveguides but are not limited to being oriented along three directional axes and accepting only $60^o$ or $120^o$ bends. They offer enhanced flexibility for the design of functional THz devices, allowing for interconnections between arbitrarily laid-out components on a chip. As a demonstration, we have used these principles to implement a four-port splitter and router. We have demonstrated a waveguide with a $90^o$ bend, as well as a more complicated waveguide with three bends and curved sections, with characteristics suitable for practical high speed data communication with a low error rate. Our demonstration represents progress towards extending the concept of topological protection to diverse non-crystalline photonic structures, including quasicrystals[48] and lattice defects[49]. This advancement holds promise for applications in on-chip THz photonics and beyond.

## Methods
### Sample fabrication
VPC interconnect chip is fabricated on a 200 $\mu m$ thick high resistivity ($\rho$) 4-inch silicon wafer with $\rho > 10\ k\Omega \cdot$ cm. In the first step, a 2 $\mu m$ thick SiO$_2$ is deposited on the top of silicon wafer. Then the photoresist AZ 5214E of thickness 1.5 $\mu m$ is coated and triangular shapes are patterned using photolithography. Reactive ion etching (RIE) is used to selectively (in the triangular region) etch SiO2, exposing the silicon surface. Deep reactive ion etching (DRIE) of the triangular region of silicon is performed to achieve the air-holes in the silicon membrane.

### Experimental section

### Vector-Network Analyser (VNA) extender-based characterization setup
Each VPC chip is characterized using a pair of vector network analyser (VNA) extension modules covering a frequency range of 300 GHz to 350 GHz. The extender modules are connected to a 24 GHz network analyser capable of measuring the full two-port scattering parameters. Each frequency extender (220-325 GHz and 325-500 GHz) has hollow-core waveguide (HCW) outputs in relation to the extender bandwidth (WR3.4 and WR2.2, respectively). The system was first waveguide calibrated using a TRM (Thru, Reflect, Match) waveguide calibration procedure. After calibration, Si tapers of VPC chip are inserted in the HCWs to inject the signal into the chip via Port 1, and transmitted signal is then collected from the respective output ports (Port 2, Port 3, Port 4). Reference (0 dB S21) is corresponding to probe-probe direct transmission (see Supplementary Information Fig. S4.1).

### THz communication setup
Data communication experiments were conducted by a photonics-based THz transmitter and electronics-based receiver, used at detection side. Two C-band continuous-wave laser lines (193.38 THz, 193.711 THz, with 331 GHz difference frequency) are used as optical carriers. One of the lines (193.38 THz) is optically amplitude/phase modulated using a Mach-Zehnder modulator (MZM). The MZM is driven by an arbitrary waveform generator (AWG) to imprint QAM-16 data signal (18 GBaud or 8 GBaud) on optical frequency. After the MZM, the two lines are combined and optically amplified using an erbium-doped fiber amplifier (EDFA). The resulting optical signal is coupled to the THz photo mixer (uni-travelling carrier photodiode, UTC-PD) for 331 GHz generation. The UTC-PD is then coupled into Port 1 of the amorphous VPC waveguide (device under test i.e., DUT) using the same hollow-waveguide-core (HWC) waveguide used during the VNA measurement. At the output Port 2, the same waveguide is used to feed the modulated THz carrier to sub-harmonic mixer (SHM, GaAs Schottky diodes technology). To extract the base band signal, a local oscillator (LO) via an electronic multiplication chain pumps this SHM. The IF signal (base band signal) at SHM output is further amplified before detection using a fast real-time oscilloscope (RTO). I/Q maps and EVM are computed using the VSA software and MATLAB routines. The detailed block diagram of the THz communication setup is presented in Supplementary Information S4.

**Numerical simulation**

The numerical results of the band structure and Berry curvature of the VPC are obtained through simulations using a finite-element method solver (COMSOL Multiphysics). The transmissions and corresponding field distributions are numerically obtained using time-domain simulations in CST Microwave Studios. The details are discussed in Supplementary Information.

**Acknowledgments**

This work was supported by the National Research Foundation (NRF), Singapore under its Competitive Research Programme NRF-CRP23-2019-0005. YC acknowledges support from NRF Singapore under NRF Investigatorship NRF-NRFI08-2022-0001. GD was partially supported by the IEMN Flagship on Ultra-High datarates (UHD). This work was also supported by the France 2030 programs, PEPR (Programmes et Equipements Prioritaires pour la Recherche) and CPER Wavetech. The PEPR is operated by the Agence Nationale de la Recherche (ANR), under the grants ANR-22-PEEL-0006 (FUNTERA, PEPR 'Electronics') and ANR-22-PEFT-0006 (NF-SYSTERA,

PEPR 5G and beyond - Future Networks). The Contrat de Plan Etat-Region (CPER) WaveTech is supported by the Ministry of Higher Education and Research, the Hauts-de-France Regional council, the Lille European Metropolis (MEL), the Institute of Physics of the French National Centre for Scientific Research (CNRS) and the European Regional Development Fund (ERDF).

**Competing Interests**

The authors declare no competing interests.

**Authors Contributions**

R.B., Y.C., and R.S. conceived the idea. R.B. designed the structures and performed the theoretical analysis. R.B. carried out the numerical simulations in CST with the help of R.J. and A.K. T.T.C, M.G., A.K., and G.D performed the experiments. P.S. and G.D. carried out the THz communication measurements. R.B., A.K., M.G., G.D., Y.C., and R.S. wrote the manuscript. Y.C. and R.S. supervised the project.

**Data availability**

The data supporting the plots within this paper are available in the data repository for Nanyang Technological University at this link (URL to be inserted later upon publication). Additional findings of this study are available from the corresponding authors upon reasonable request.

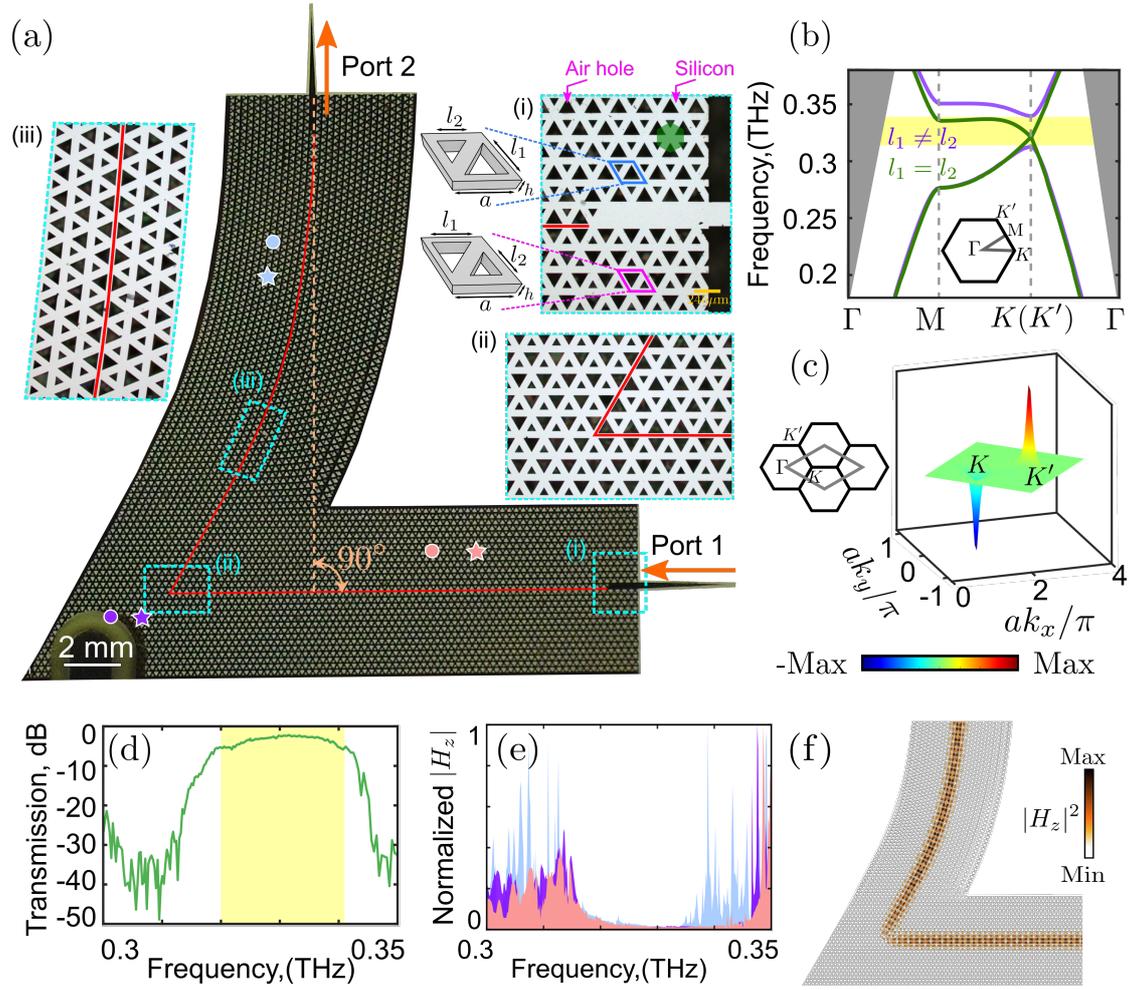

**Figure 1 | On-chip THz amorphous VPC waveguide and topology. a,** Optical image of the fabricated sample, which channels the input signal (port 1) to a perpendicular output port (port 2). The interface between the two domains is indicated by the red line. **b,** Band structures for the TE modes without and with broken inversion symmetry, drawn in green and violet respectively. For the former case, $l_1 = l_2 = 0.5a_0$; for the latter case, $l_1 = 0.65a_0$ and $l_2 = 0.35a_0$, where $a_0 = 242.5\,\mu$m. The undeformed VPC has periodicity $a = 245\,\mu$m, and the thickness of the silicon membrane is $h = 200\,\mu$m. The grey area represents the region above the light line. **c,** Berry curvature for the lowest band of the unit cell indicated in blue in **a**. **d,** Experimentally measured transmission spectra, showing near unity transmission over 0.32-0.341 THz (yellow shaded region). **e,** Frequency-dependent normalized field distribution obtained from numerical simulations, for probes placed at the filled circles and sources placed at stars with matching colors in subplot **a**. The reduced bulk response around 0.33 THz is consistent with the formation of a bulk bandgap. **f,** Simulated field intensity at 0.328 THz.

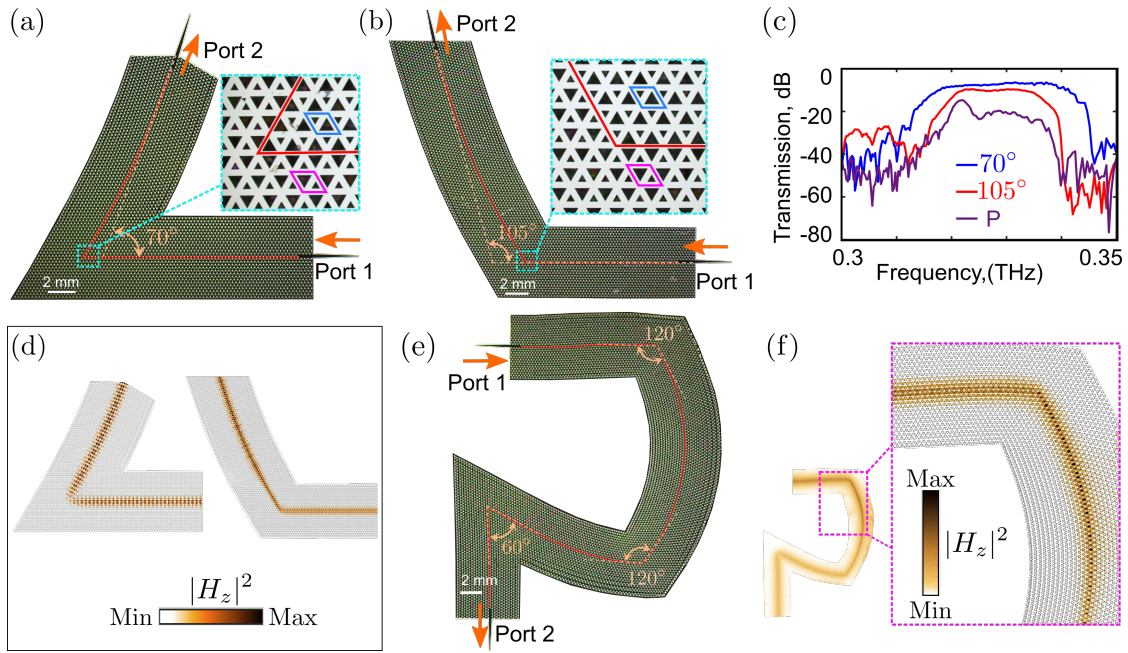

**Figure 2 | THz VPC waveguide with arbitrary shape. a-b,** Optical image of the fabricated samples, which direct the input signal to output ports positioned at $70^o$ and $105^o$, respectively. **c,** Experimentally recorded transmission spectra for these two waveguides, as well as the structure in **e**. **d,** Simulated intensity distributions for the two samples at 0.326 THz and 0.324 THz. **e,** Optical image of a more complex VPC structure, featuring a P-shaped waveguide. **f,** Simulated field distribution for the P-shaped waveguide at 0.32 THz, including a zoomed-in view near one of the bends. The experimentally measured transmission spectrum is shown in subplot **c**.

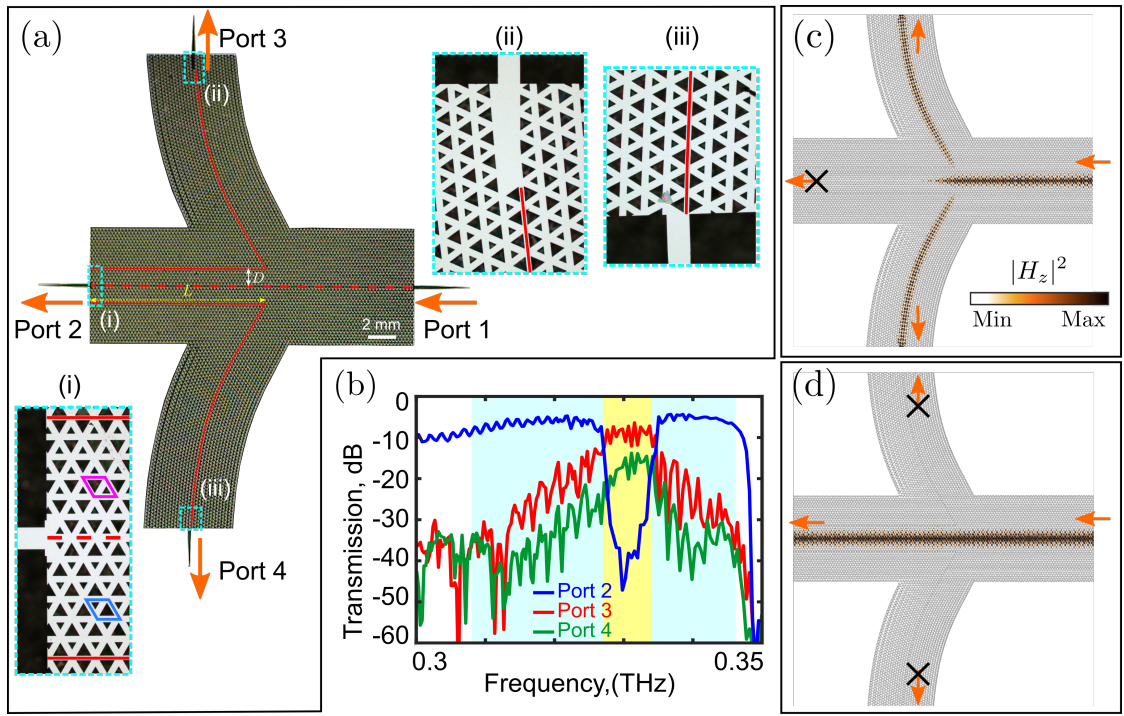

**Figure 3 | On-chip THz functional re-router and splitter. a,** Optical image of the device. **b,** Experimentally measured transmission spectra at ports 2, 3, and 4. The input signal at port 1 is routed to port 2 in the frequency windows shaded in cyan (0.308-0.327 THz and 0.334-0.346 THz). In the yellow-shaded window (0.327-0.334 THz), the input is divided into two perpendicular output ports 3 and 4. **c,** Simulated intensity distribution at 0.328 THz, within the yellow-shaded frequency window. **d,** Similar plot as in **c**, but at 0.335 THz, within the cyan-shaded frequency window.

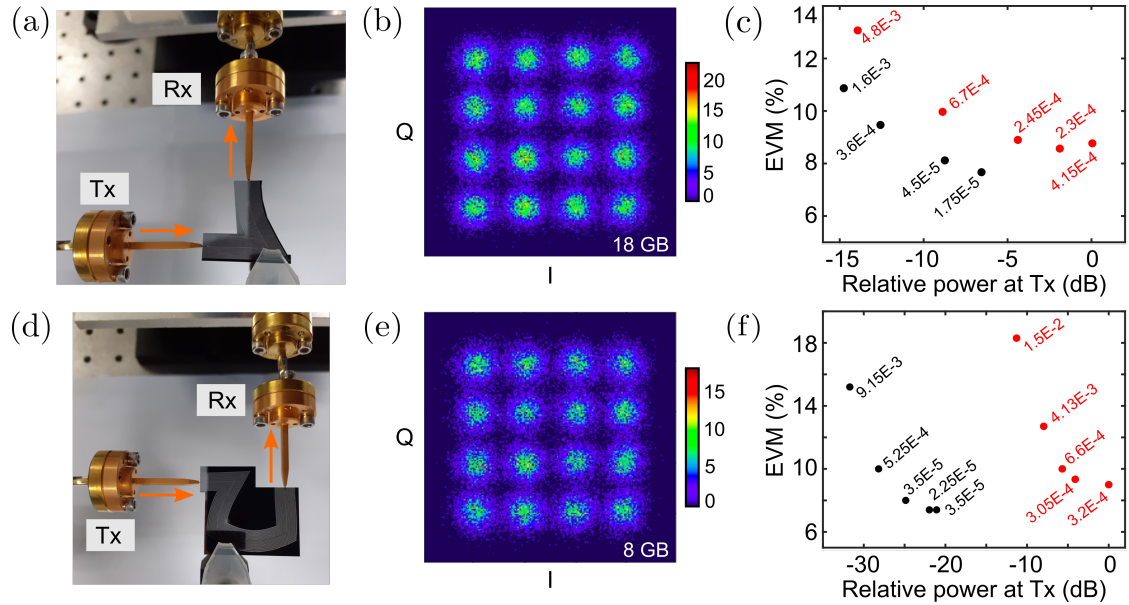

**Figure 4 | High speed THz communication. a,d,** Photographs of the experimental setup with $90^o$ and P-shaped VPC chips. **b,e,** I-Q constellation diagram at 18 GB (72 Gbit/s) and 8 GB (32 Gbit/s), respectively, recorded using QAM-16 modulation format. **c,f,** EVM versus relative transmitted power with (red dots) and without (black dots) the VPC chips. Each data point is labelled with its corresponding bit-error-rate (BER).